\documentclass[11pt,twoside]{article}

\usepackage{asp2021}

\aspSuppressVolSlug
\resetcounters

\begin{document}

\title{Role of General Users in the Lifecycle of Scientific Software}

\author{Bjorn Emonts,$^{1,2}$ and the CASA Team}
\affil{$^1$National Radio Astronomy Observatory, 520 Edgemont Road, Charlottesville, VA 22903, USA; \email{bemonts@nrao.edu}}
\affil{$^2$CASA User Liaison; \email{casa-feedback@nrao.edu}}
\paperauthor{Bjorn Emonts}{bemonts@nrao.edu}{0000-0003-2983-815X}{National Radio Astronomy Observatory}{520 Edgemont Road}{Charlottesville}{VA}{22903}{USA}



\begin{abstract}
In science, the lifecycle of software products is typically managed with limited resources while facing unlimited demand. Scientific software requirements are necessarily often dominated by internal project specifications and deadlines, but these internal priorities, while beneficial for the community as a whole, do not always align with the individual needs of our ultimate customers: general users. For software products to have the broadest reach, ideally the general user community should be involved in all aspects of the data lifecycle, but reality is that user expectations need to be managed. Based on the lifecycle of the Common Astronomy Software Applications for radio astronomy (CASA), we will show avenues for software teams to interact with general users, even when facing limited resources for user support. We will discuss how involvement of users and user groups in prioritizing software development can benefit both the user community and the software teams. The contents of these proceedings were presented at the 35th conference on Astronomical Data Analysis Software $\&$ Systems (ADASS XXXV).
\end{abstract}



\section{Introduction}

Astronomy is a field of research with a large scope and reach \citep{ASTRO2020}. It relies on state-of-the-art observatories and world-class computing facilities, as well as the astronomical software packages that are needed to process and analyze data from these facilities. An historic overview of the widespread introduction of software in astronomy was published by \citet{sho01}. Astronomical software is often highly specialistic with a limited user-base, consisting of mostly scientists and software engineers. This puts limits on the resources that are available for the development, maintenance, and support of astronomical software.  

The Common Astronomy Software Applications (CASA) is a versatile software package for the processing of radio astronomical data from various radio observatories \citep{casa22}. It has a wide user-base among the community, and is managed by an international development team. In these proceedings, we describe the role that general users play in the lifecycle of the CASA software. Section \ref{sec:CASA} gives a high-level overview of CASA and its development lifecycle. Section \ref{sec:role} highlights the role that general users play in the development  lifecycle, and how stakeholder representation provides critical input for the long-lived health and versatility of the software. In Sect. \ref{sec:support}, we address how the CASA team balances user support with available resources. Section \ref{sec:conclusions} concludes with a brief outlook on the future of CASA, while Sect.\,\ref{sec:resources} provides an overview of CASA resources that are available to the user community.

\section{The CASA Software for Radio Astronomy}
\label{sec:CASA}

CASA is the {\sl Common Astronomy Software Applications}, which is an open-source software package for radio astronomy \citep{casa22}. CASA is the official software for the Karl G. Jansky Very Large Array (VLA), Atacama Large Millimeter/submillimeter Array (ALMA), and Nobeyama 45m Telescope, but is often used also for the calibration and imaging of data from other radio telescopes. CASA can process interferometric, single-dish, and Very Long Baseline Interferometry (VLBI) data.

CASA can be run interactively and through scripting, and a core functionality is to support calibration and imaging pipelines for ALMA, VLA, and Nobeyama \citep{hun23}. CASA is implemented in C++ and accessible through IPython \citep{per07}. It is built on top
of Casacore, which contains the core libraries, and which is a nearly static platform that is used by many radio observatories \citep{casacore19}. The CASA software consists of the following core functionality:
\begin{itemize}
\item{{\sl Tools:} basic C++ functions as Python-class objects with {\sl methods} that perform basic operations on the data;} 
\item{{\sl Tasks:} user-friendly bundles of tools and Python functionality that perform specific steps in the data processing. CASA tasks contain {\sl parameters} that give users flexibility to control the calibration and imaging;} 
\item{{\sl GUIs:} Graphical User Interfaces (GUIs) to visualize and examine data and images;} 
\item{{\sl Data Repository:} external data that changes over time, and which is needed for accurate calibration and imaging. This includes reference frames, Earth Orientation parameters, and telescope-specific models.}
\end{itemize}

\noindent Historically, CASA has been offered as an all-inclusive package, but since CASA 6 the software has been available in various formats:
\begin{itemize}
\item{{\sl Monolithic tar-file:}} this is the all-inclusive, 'plug-and-play' format that contains a Python interpreter and all necessary libraries;
\item{{\sl Modular pip-wheels:}} this format allows users to run CASA tasks, tools, and GUIs in their own Python environment, or in Jupyter Notebooks or Google Colab;
\item{{\sl Pipeline version:}} certain tar-file versions of CASA are packaged together with the calibration and imaging pipelines for ALMA and the VLA.
\end{itemize}

CASA is being developed by an international team of around 30 scientists and software engineers based at the National Radio Astronomy Observatory (NRAO), the European Southern Observatory (ESO), the National Astronomical Observatory of Japan (NAOJ), and the Joint Institute for Very Long Baseline Interferometry European Research Consortium (JIVE), under the guidance of NRAO. 

\subsection{Lifecycle of CASA}
\label{sec:lifecycle}

The CASA team strives to maintain a "rolling" release schedule, with a new CASA release roughly every three months (Fig.\,\ref{fig:lifecycle}). Such a rolling release schedule simplifies and standardizes the development cycle and release process. This quarterly release is available both in monolithic and modular format (see Sect.\,\ref{sec:CASA}). Once per year, the interferometric pipeline that supports ALMA, VLA, and the VLA Sky Survey (VLASS) is bundled with CASA, and released in monolithic tar-file format. While ALMA, the VLA, and VLASS strive to use the same pipeline, there can be small differences that may lead to different flavors of the pipeline, each with its own CASA release version.

Development priorities are set by CASA management following input from a large number of stakeholders, who provide requirements and feature requests twice a year. This is complemented with a yearly report from the DMS Panel of the Users Committee, which is an advisory committee on the development of software at NRAO, which operates on behalf of the general user community (see Sect.\,\ref{sec:stakeholder}). 

\articlefigure[width=0.9\textwidth]{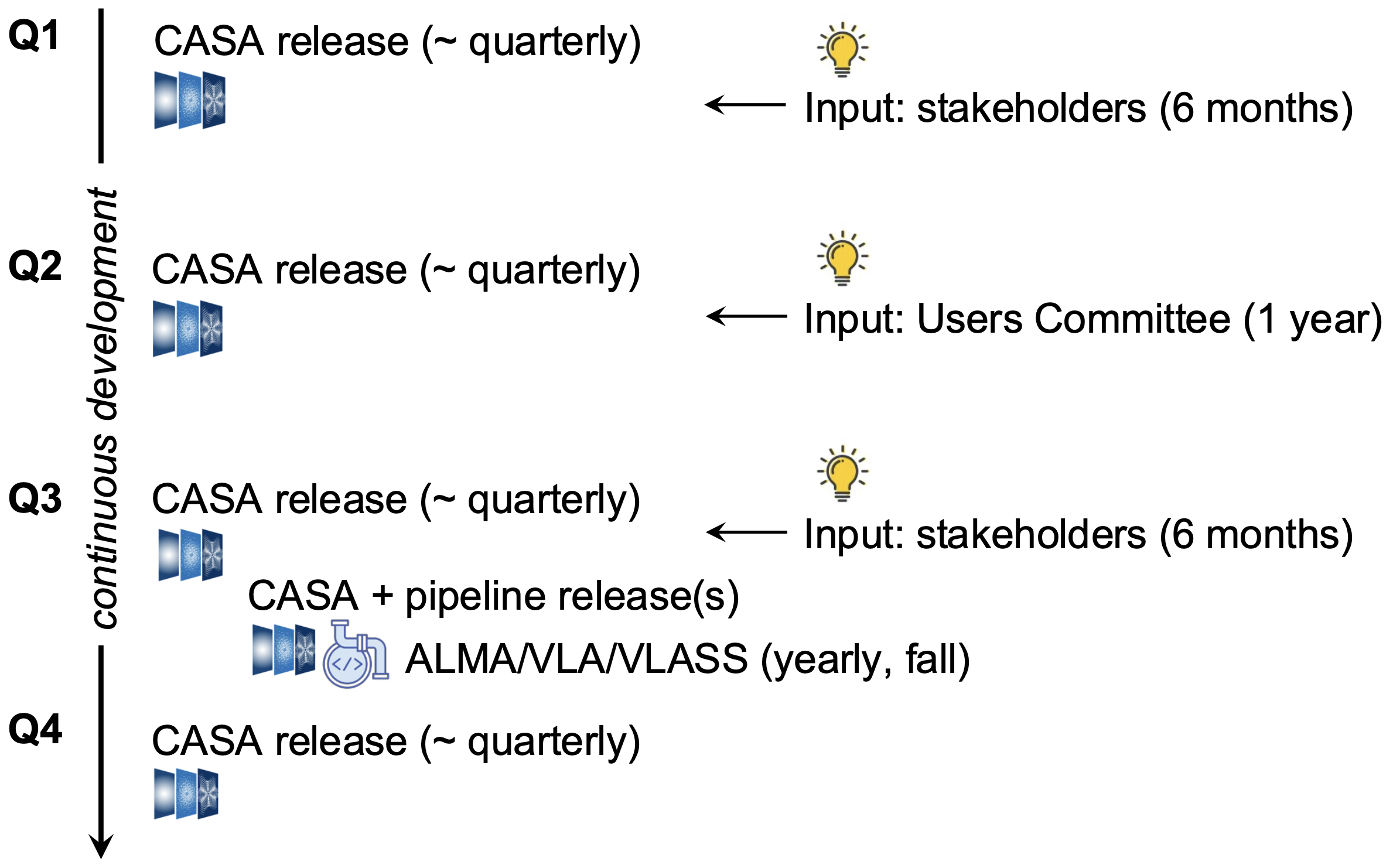}{fig:lifecycle}{Lifecycle of the CASA software. CASA is released quarterly, with once a year a CASA release bundled with the pipeline for ALMA, VLA, and VLASS. Stakeholder input is gathered every six months, while once a year the DMS Panel of the Users Committee (DPUC) provides a report with recommendations on behalf of the user community.}

\section{Role of Users in Software Development}
\label{sec:role}

General users are the ultimate customers of software products. As such, they provide unique and important input critical for the long-terms health and versatility of the software. This is particularly true if the software is intended for broad use, as is the case for CASA. The problem is that general users form a silent majority that is not as vocal as institutional stakeholders and experts, or power-users. It is good to keep in mind that general users often see scientific software as a means-to-an-end, not the end goal.

A solution to this problem is to capture feedback from the general user community. This can be done passively, for example by collecting software statistics, telemetry data, or crash reports, or actively, through user surveys, review panels, or direct contact (e.g., helpdesk or email). In the case of CASA, we have given the user community a voice as stakeholder, which has resulted in unique insights into essential requirements that might otherwise not have been obtained.

\subsection{Users as CASA Stakeholders}
\label{sec:stakeholder}

As shown in Fig.\,\ref{fig:stakeholder}, CASA has a large number of stakeholders. Stakeholders whom we identify in these proceedings as "internal" are ALMA, VLA, VLBA, and groups at NRAO's department of Data Management and Software (DMS). These internal stakeholders are represented by local colleagues, and therefore have a strong voice to advocate institutional requirements. These requirements are often short-term priorities essential for telescope operations, such as specific tasks to support new observing modes, new parameters or defaults for pipeline operations, or bug that block data processing. 

It takes a more active approach to obtain feedback from the general users. For CASA, the user community also has a voice as stakeholder, which we here identify as "external" stakeholder, with representation through both the CASA User Liaison and the DMS Panel of the Users Committee (DPUC). The CASA User Liaison (author of these proceedings) is a member of the CASA team, first contact for outside users, and stakeholder representative on behalf of external users. The DPUC is an software sub-committee of the NRAO Users Committee, and consist of ten user representatives that advise NRAO's devision on Data Management and Systems (DMS) on all things related to software via a yearly DPUC meeting and written report with recommendations. 

\articlefigure[width=0.9\textwidth]{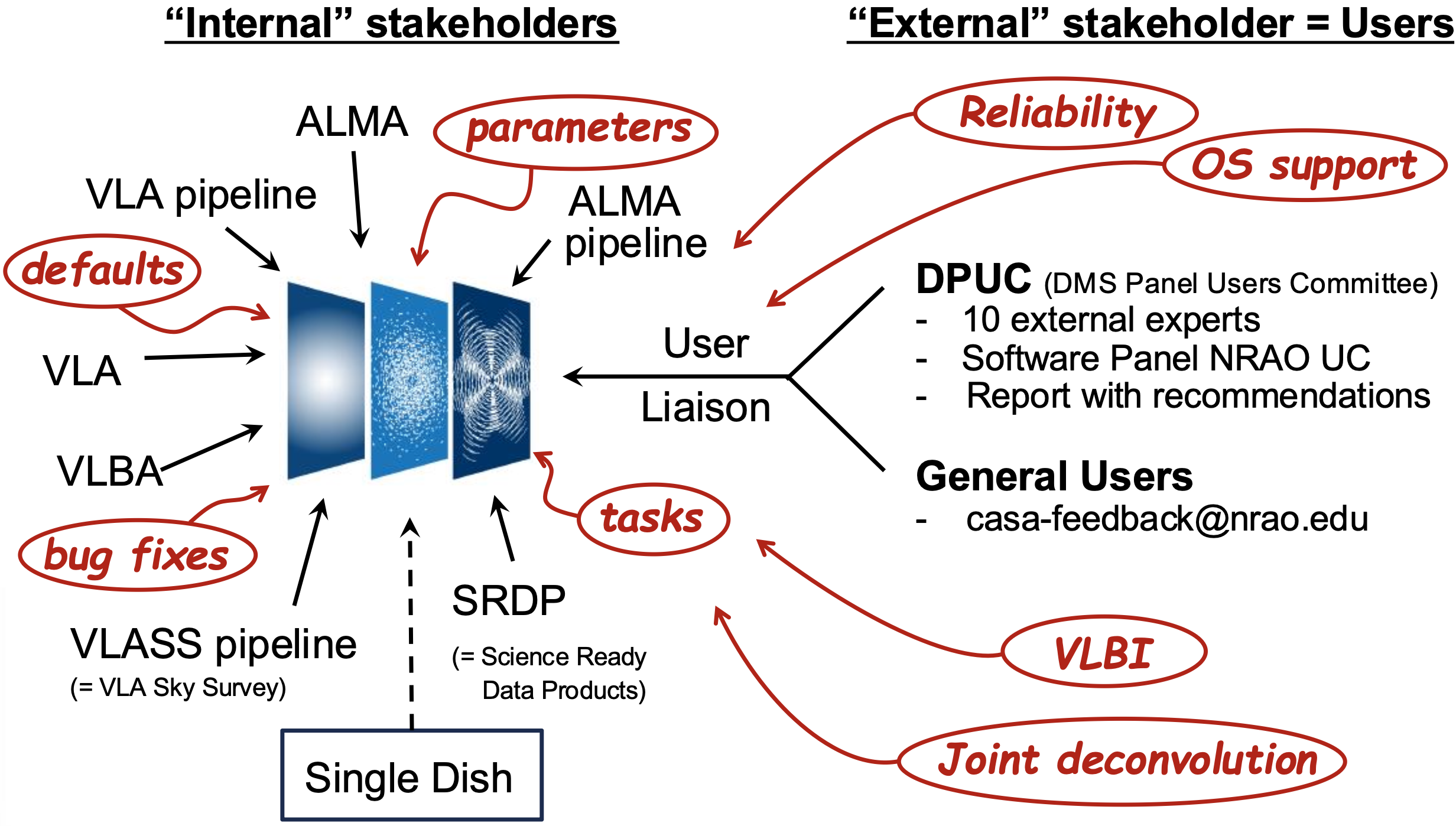}{fig:stakeholder}{CASA development priorities through stakeholder representation. "Internal" stakeholders represent ALMA, VLA, VLBA, and other groups at NRAO, with Single Dish priorities handled by NAOJ. General users have "external" stakeholder representation through the CASA User Liaison and DMS Panel of the Users Committee. Internal stakeholders often focus on short-term requirements needed for operations, while the user community provides input on long-term requirements that take a `big-picture' approach to keeping CASA versatile and robust.}

Experience shows that requirements from the external user community focus on longer term goals that tackle the "big-picture" approach to keeping CASA a versatile and robust software package. An example has been the desire by the community to make CASA more reliable \citep{memo6}, which triggered the CASA team to implement an extensive and robust testing philosophy and infrastructure (see contribution of S. Castro as part of these proceedings). Other examples are the continuous request to keep support for Mac OS (not used in telescope operations), increased support for VLBI, and the request to offer the possibility for joint deconvolution of single-dish and interferometry data (which resulted in the development a new task \texttt{sdintimaging}). The general user community therefore provides a critical role in the healthy, long-term development of the CASA software.

\section{User Support}
\label{sec:support}

Development teams also need to provide feedback to users regarding the use of their software, and assist when problems arise. User support requires a dedicated investment of a certain full- or fractional-time equivalent (FTE) workload. It is key that software teams balance user support based on available FTEs, to avoid distractions among the development team, and keep focus on maintaining and improving the software. A main challenge in this is to manage user expectations accordingly.

Software teams have many avenues to provide a balanced user support. These include {\sl active support} in the form of personal contact between the software team and users, sometimes via formal platforms, such as helpdesks. The user community may also benefit from {\sl passive support} in the form of tutorials and demos, Newsletters, or an online forum. A third avenue that is essential support for users, and which is almost mandatory for any software team, is what we define here as {\sl technical support}, which includes code documentation, as well as a dedicated suite of tests that run on various Operating Systems (OSs). In the next Sections, we will address how CASA provides some of these critical support structures.

\subsection{CASA Feedback: casa-feedback@nrao.edu}
\label{sec:feedback}

The need for balancing user support with resources is often most critical regarding one-on-one user support. A dedicated helpdesk, where users can raise problems and expect swift help to find a resolution, is a platform that is most useful for the user community, but can become a serious constraint for development teams. This is visualized in Fig.\,\ref{fig:helpdesk}, where we summarize CASA-specific tickets that where handled by the telescope support teams through the NRAO and ALMA Helpdesk during the period of ten months \citep{memo6}. The majority of incoming tickets were related to misunderstandings by the user, bad data, or issues that had been resolved in more recent CASA releases. Only a minority of tickets revealed a problem with the CASA code or documentation. While it is critical to capture this feedback from the community to ensure that problems with the code are addressed, the workload involved in sorting out software problems from the non-software issues can be significant. Therefore, running an official helpdesk can diminish the often already strained resources of software teams for code development.

\articlefigure[width=0.85\textwidth]{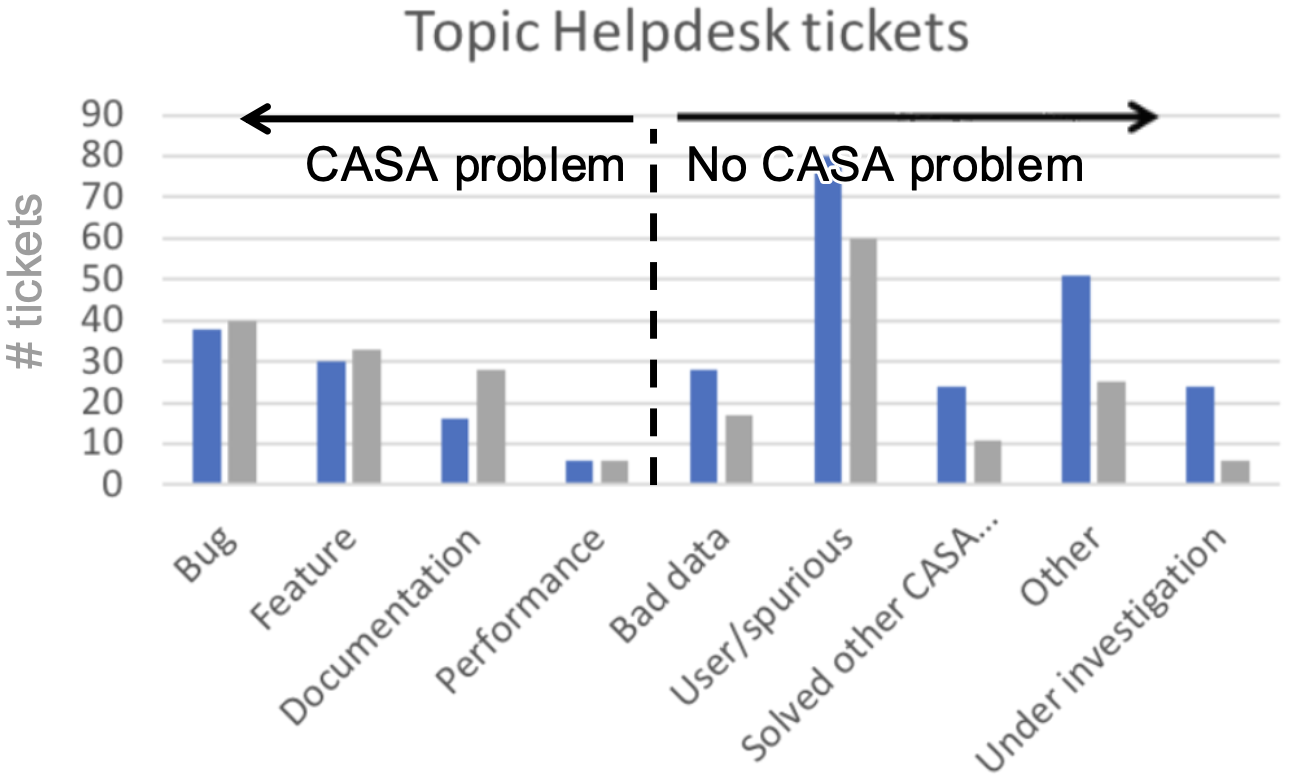}{fig:helpdesk}{Helpdesk statistics. Total number of CASA-labeled Helpdesk tickets received by the NRAO and ALMA helpdesks during a 10-month period in 2018, grouped by topic. Only a minority of tickets revealed a problem with the CASA code base or documentation. Reproduced from CASA Memo 6 \citep{memo6}.}

A more informal avenue for communication with the user community may be beneficial if FTEs for user support are limited. The CASA team adopted such an approached by introducing a dedicated contact email: casa-feedback@nrao.edu. Casa-feedback@nrao.edu is intended for general feedback to the CASA team, and is also used for best-effort support for users who encounter problems with the CASA code or documentation. Contrary to a helpdesk system, this dedicated email address does not guarantee one-on-one problem solving or automated archiving of tickets, but it has several advantages that keep the process efficient and manageable:
\begin{itemize}
\item{{\sl Best-effort:} a best-effort support system minimizes the pressure on development resources, allowing developers to respond to reported issues when the development schedule allows;}
\item{{\sl Direct CASA contact:} the CASA team is not actively involved on the NRAO and ALMA helpdesks, therefore a dedicated email provides a direct means for users (and helpdesk staff) to contact the CASA team;}
\item{{\sl Multiple recipients:} a dedicated email system allows multiple recipients, which means that we have multiple sets of eyes on incoming issues. This increases the response time while minimizing the overheads involved in consulting other team-members;}
\item{{\sl Feedback:} casa-feedback@nrao.edu is also an easy avenue for users to provide general feedback on the CASA software, and on its successor, the Radio Astronomy Data Processing System (RADPS; see Sect.\,\ref{sec:conclusions}).}
\end{itemize}
We welcome users to use casa-feedback@nrao.edu for feedback or questions concerning CASA.

\subsection{CASA Documentation}
\label{sec:docs}

Code documentation is often the first point of contact for users with questions about the software. As such, it is essential that software packages include a complete and accurate documentation that users can read, and be referred to, when encountering problems.

CASA Docs\footnote{\url{https://casadocs.readthedocs.io}} is the official documentation of the CASA sofware (Fig.\,\ref{fig:casadocs}). It is written alongside development of the codebase. In addition, tests to verify that the CASA code is working as intended are  written against CASA Docs. This optimizes efforts to ensures that the codebase and documentation are in sync and accurate.

CASA Docs provides extensive descriptions of CASA tasks and their parameters. These task pages can optionally be opened from the CASA command-line by typing \texttt{doc('taskname')}. CASA Docs also provides documentation on the tools and GUIs, as well as instructions on how to update the data repository. In addition, it contains general chapter pages with detailed background information on calibration and imaging, Release Information and Known Issues, installation instructions and OS compatibility, CASA Memos and Knowledgebase articles, Community Examples on running CASA in Jupyter Notebooks and Colab, and contact information. A new version of CASA Docs is published with each CASA release. 

The CASA software is also described in several recent publications \citep{casa22,bem22,emo25}.

\articlefigure[width=0.9\textwidth]{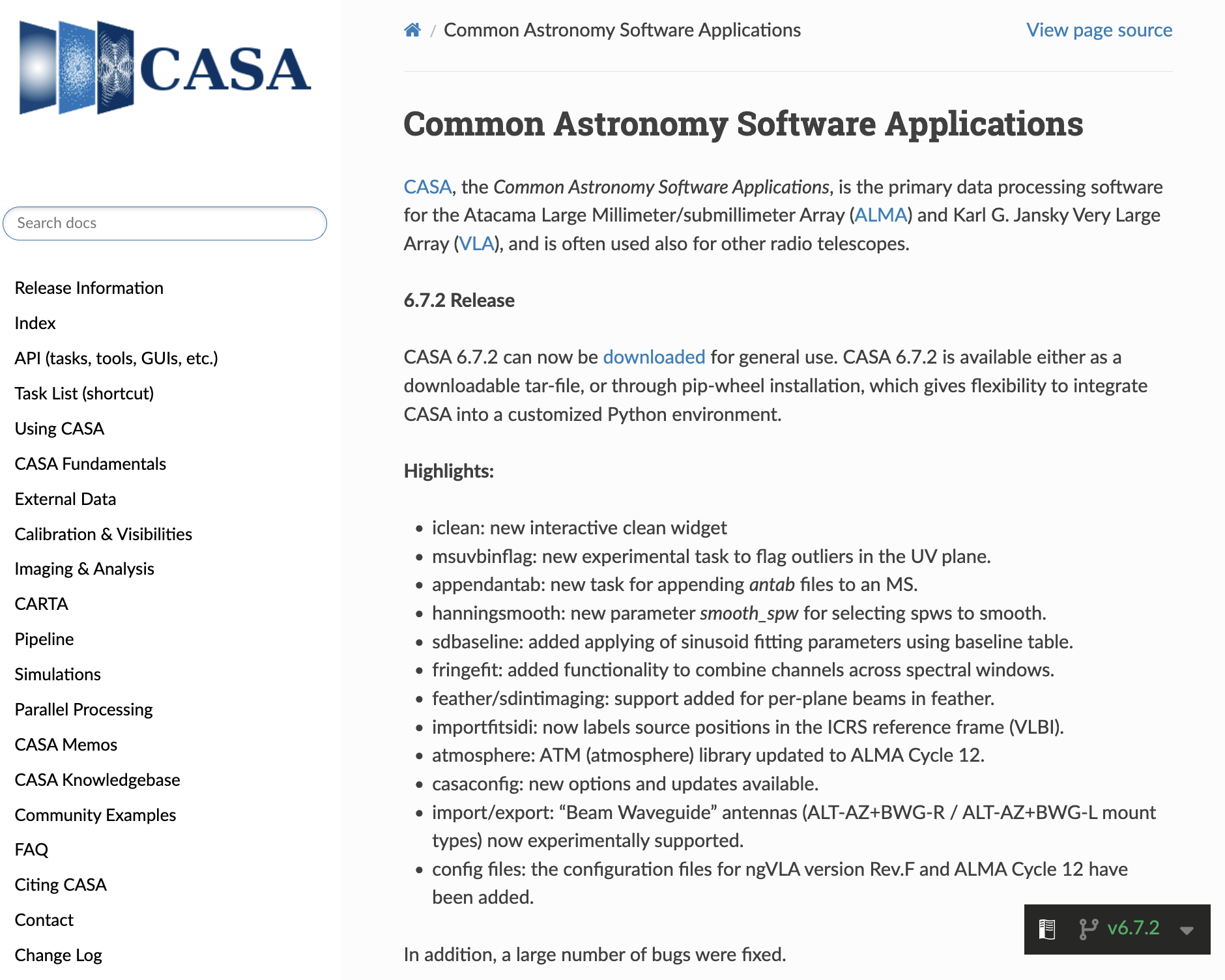}{fig:casadocs}{Layout of the CASA Docs documentation. Shown is the landing page for the 6.7.2 release version, with a navigatable list of pages and search box on the left. CASA Docs is hosted on Github and can be accesses through readthedocs: \url{https://casadocs.readthedocs.io}}

\subsection{CASA Testing}
\label{sec:tests}

A large fraction of user questions concerns CASA installation and OS support. Results from a 2018 CASA User Survey revealed that Ubuntu is the most widely used OS for running CASA among the external user community (Fig.\,\ref{fig:os}a ; \citealt{memo6}). This despite the fact that CASA is only officially supported for the latest releases of Linux RedHat and Mac OS, and only fully validated against the current operational Linux/RedHat configuration of NRAO instruments.

In order to meet user demand regarding the versatility of operating systems, the CASA team incorporated combinations of the most widely used OS and Python versions at various places in the suite of tests that validate the CASA code. The compatibility of CASA with these OS/Python versions is advertised in CASA Docs (Fig.\,\ref{fig:os}b). This approach allows the CASA team to focus efforts on developing the code against the OS/Python version used in ALMA and VLA operations, while providing feedback to the community for which other OS/Python versions CASA is expected to work. Users can report a bug through casa-feedback@nrao.edu if CASA fails on any of the compatible OS/Python versions that are advertised in CASA Docs.

\articlefigure[width=\textwidth]{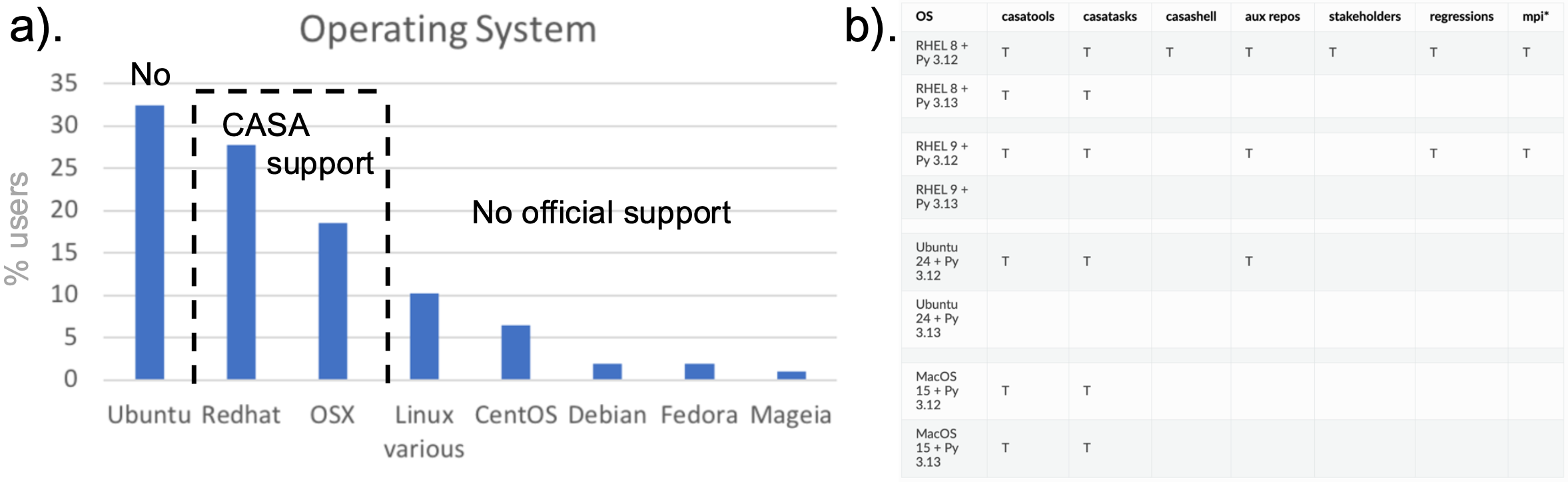}{fig:os}{CASA compatibility with different Operating Systems. {\sl a).} Survey statistics on the use of Operating Systems among the external user community. The figure is reproduced from CASA Memo 6 \citep{memo6}. The majority of users relies on Ubuntu for CASA processing, despite the fact that CASA is only officially supported for Linux/RedHat and Mac OS. {\sl b).} Automated testing table in CASA Docs, where "T" indicates tests that have run and successfully completed. This table is guidance for the community in that it lists all combinations of OS and Python that are expected to be compatible with the corresponding CASA release version.}

\section{Outlook}
\label{sec:conclusions}

CASA is a widely used data-processing software for radio astronomy, with an active role for the user community in defining development requirements, and with efficient use of resources for providing support to the community. While CASA has become the predominant software for radio astronomy, the advancement of radio telescopes, and associated dramatic increase in data rates and volumes, introduces limitations for a code that is two decades old \citep{mul07}.

For the foreseeable future, two major infrastructure projects dominate the requirements for radio data processing at NRAO and partner institutes. These are the ALMA Wideband Sensitivity Upgrade (WSU; \citealt{car23}) and the Next-Generation Very Large Array (ngVLA; \citealt{mur18}). Both instruments will provide major challenges for data processing software, resulting from orders of magnitude higher data rates and volumes \citep[e.g.,][]{bha21,kep24}. The current CASA software will not be sufficient for efficiently processing data from these instruments. 

As a result, efforts are underway to develop a next-generation Radio Astronomy Data Processing System (RADPS). RADPS will have as primary objective the production of high-level data products from ALMA-WSU and ngVLA, with secondary goal to support the evolution of radio astronomy data processing through a widely accessible package. RADPS will rely on scalable data-processing capabilities using mostly off-the-shelf technology, with intuitive user interfaces.

A detailed discussion of RADPS is beyond the scope of these proceedings. However, RADPS development heavily leans on CASA developers. It is therefore anticipated that new development within CASA will continue to wind down in favor of RADPS. This will result in CASA transitioning towards a maintenance mode in the foreseeable future. Regardless, the legacy of CASA will be one where users played a critical role over the lifetime of the software.

\section{CASA Resources}
\label{sec:resources}

The following is an overview of CASA resources available to the user community:
\begin{itemize}
\item{{\bf CASA contact:} we welcome feedback from users!\\
(casa-feedback@nrao.edu)}
\item{{\bf CASA announcements:} stay up-to-date on CASA announcements and register!\\
(\url{https://listmgr.nrao.edu/mailman/listinfo/casa-announce})}
\item{{\bf CASA Website:} official CASA website\\ (\url{https://casa.nrao.edu})}
\item{{\bf CASA Docs:} official CASA documentation, or {\sl "How the code works"}\\ 
(\url{https://casadocs.readthedocs.io})}
\item{{\bf CASA Guides:} data reduction tutorials, or {\sl "How to work the code"}\\
(\url{https://casaguides.nrao.edu} -- maintained by instrument support teams)}
\item{{\bf CASA Code:} open-source codebase on Github and Bitbucket\\
(\url{https://github.com/casangi})\\
(\url{https://open-bitbucket.nrao.edu/projects/CASA/repos/casa6/browse})}
\item{{\bf CASA Newsletter:} CASA news sent to the community\\
(\url{https://science.nrao.edu/enews/casa_012/index.shtml})}
\item{{\bf CASA Reference Paper:} CASA Team et al. 2022, PASP, 134, 114501\\
(\url{https://iopscience.iop.org/article/10.1088/1538-3873/ac9642})}
\item{{\bf DMS Panel of the Users Committee (DPUC):} software advisory committee\\ 
(\url{https://safe.nrao.edu/wiki/bin/view/Software/CASA/CASAUsersCommittee})}
\end{itemize}

\acknowledgements 

We thank the organizers of ADASS XXXV for a great conference. We are indebted to the current and past members of the DMS Panel of the Users Committee (formerly the CASA Users Committee), the user community, and the CASA stakeholders for continuously providing invaluable feedback that drives CASA development. We also thank our colleagues in the pipeline development teams and instrument support groups for their valuable efforts regarding code validation, quality control, and user support. CASA is being developed by an international team of scientists and software engineers based at the National Radio Astronomy Observatory (NRAO), the European Southern Observatory (ESO), the National Astronomical Observatory of Japan (NAOJ), and the Joint Institute for Very Long Baseline Interferometry European Research Consortium (JIVE), under the guidance of NRAO. The National Radio Astronomy Observatory is a facility of the National Science Foundation operated under cooperative agreement by Associated Universities, Inc. ALMA is a partnership of ESO (representing its member states), NSF (USA) and NINS (Japan), together with NRC (Canada), MOST and ASIAA (Taiwan), and KASI (Republic of Korea), in cooperation with the Republic of Chile. The Joint ALMA Observatory is operated by ESO, AUI/NRAO and NAOJ. 

\bibliography{043.bib}  

@MISC{bha21,
       author = {{Bhatnagar}, S. and {Hiriart}, R. and {Pokorny}, M.},
        title = "{Size-of-Computing Estimates for ngVLA Synthesis Imaging}",
 howpublished = {ngVLA Computing Memo $\#$4, August 2021, 22 pages},
         year = 2021,
        month = aug,
        pages = {},
       adsurl = {https://ngvla.nrao.edu/page/memos#comp-memo}
}

@INPROCEEDINGS{mul07,
       author = {{McMullin}, J.~P. and {Waters}, B. and {Schiebel}, D. and {Young}, W. and {Golap}, K.},
        title = "{CASA Architecture and Applications}",
    booktitle = {Astronomical Data Analysis Software and Systems XVI},
         year = 2007,
       editor = {{Shaw}, R.~A. and {Hill}, F. and {Bell}, D.~J.},
       series = {Astronomical Society of the Pacific Conference Series},
       volume = {376},
        month = oct,
        pages = {127},
       adsurl = {https://ui.adsabs.harvard.edu/abs/2007ASPC..376..127M}
}

@PROCEEDINGS{mur18,
        title = "{Science with a Next Generation Very Large Array}",
    booktitle = {Science with a Next Generation Very Large Array},
         year = 2018,
       editor = {{Murphy}, Eric},
       series = {Astronomical Society of the Pacific Conference Series},
       volume = {517},
        month = dec,
       adsurl = {https://ui.adsabs.harvard.edu/abs/2018ASPC..517.....M}
}

@MISC{kep24,
       author = {{Kepley}, Amanda A. and {Brogan}, Crystal L. and {Carpenter}, John and {Diaz Trigo}, Maria and {Hatsukade}, Bunyo and {Antognini}, Jonathan},
        title = "{Estimates of ALMA WSU Data Properties}",
 howpublished = {ALMA Memo Series, 626, January 31, 2024, 40 pages},
         year = 2024,
        month = jan,
        pages = {E1},
       adsurl = {https://ui.adsabs.harvard.edu/abs/2024alma.reptE...1K}
}

@INPROCEEDINGS{car23,
       author = {{Carpenter}, John and {Brogan}, Crystal and {Iono}, Daisuke and {Mroczkowski}, Tony},
        title = "{The ALMA Wideband Sensitivity Upgrade}",
    booktitle = {Physics and Chemistry of Star Formation: The Dynamical ISM Across Time and Spatial Scales},
         year = 2023,
       editor = {{Ossenkopf-Okada}, V. and {Schaaf}, R. and {Breloy}, I. and {Stutzki}, J.},
        month = feb,
        pages = {304},
          doi = {10.48550/arXiv.2211.00195},
archivePrefix = {arXiv},
       eprint = {2211.00195},
 primaryClass = {astro-ph.IM},
       adsurl = {https://ui.adsabs.harvard.edu/abs/2023pcsf.conf..304C}
}

@ARTICLE{bem22,
       author = {{van Bemmel}, Ilse M. and others},
        title = "{CASA on the Fringe-Development of VLBI Processing Capabilities for CASA}",
      journal = {\pasp},
         year = 2022,
        month = nov,
       volume = {134},
       number = {1041},
          eid = {114502},
        pages = {114502},
          doi = {10.1088/1538-3873/ac81ed},
archivePrefix = {arXiv},
       eprint = {2210.02275},
 primaryClass = {astro-ph.IM},
       adsurl = {https://ui.adsabs.harvard.edu/abs/2022PASP..134k4502V}
}

@INPROCEEDINGS{emo25,
       author = {{Emonts}, Bjorn and {van Bemmel}, Ilse and {Moellenbrock}, George and {Kettenis}, Mark and {Small}, Des and {Rau}, Urvashi and {the CASA Team} and {the CASA-VLBI Team}},
        title = "{CASA, the Common Astronomy Software Applications for Radio Astronomy and VLBI: a Reference Summary}",
    booktitle = {Astronomical Data Analysis Software and Systems XXXII},
         year = 2025,
       editor = {{Gaudet}, S{\'e}verin and {Bohlender}, David and {Gwyn}, Stephen and {Hincks}, Adam and {Teuben}, Peter},
       series = {Astronomical Society of the Pacific Conference Series},
       volume = {538},
        month = sep,
        pages = {154},
          doi = {10.26624/WTOV6553},
       adsurl = {https://ui.adsabs.harvard.edu/abs/2025ASPC..538..154E}
}

@MISC{memo6,
       author = {{Emonts}},
        title = "{CASA Memo 6: User Survey and Helpdesk Statistics}",
 howpublished = {CASA Docs Memo Series},
         year = 2018,
        month = Nov,
          eid = {},
archivePrefix = {},
       eprint = {},
       url = {https://casadocs.readthedocs.io/en/stable/notebooks/memo-series.html}
}

@ARTICLE{per07,
       author = {{Perez}, Fernando and {Granger}, Brian E.},
        title = "{IPython: A System for Interactive Scientific Computing}",
      journal = {Computing in Science and Engineering},
         year = 2007,
        month = jan,
       volume = {9},
       number = {3},
        pages = {21-29},
          doi = {10.1109/MCSE.2007.53},
       adsurl = {https://ui.adsabs.harvard.edu/abs/2007CSE.....9c..21P}
}

@MISC{casacore19,
       author = {{Casacore Team}},
        title = "{casacore: Suite of C++ libraries for radio astronomy data processing}",
 howpublished = {Astrophysics Source Code Library, record ascl:1912.002},
         year = 2019,
        month = dec,
          eid = {ascl:1912.002},
archivePrefix = {ascl},
       eprint = {1912.002},
       adsurl = {https://ui.adsabs.harvard.edu/abs/2019ascl.soft12002C}
}

@ARTICLE{hun23,
       author = {{Hunter}, Todd. R and others},
        title = "{The ALMA Interferometric Pipeline Heuristics}",
      journal = {\pasp},
         year = 2023,
        month = jul,
       volume = {135},
       number = {1049},
          eid = {074501},
        pages = {074501},
          doi = {10.1088/1538-3873/ace216},
archivePrefix = {arXiv},
       eprint = {2306.07420},
 primaryClass = {astro-ph.IM},
       adsurl = {https://ui.adsabs.harvard.edu/abs/2023PASP..135g4501H}
}

@ARTICLE{casa22,
       author = {{CASA Team} and others},
        title = "{CASA, the Common Astronomy Software Applications for Radio Astronomy}",
      journal = {\pasp},
         year = 2022,
        month = nov,
       volume = {134},
       number = {1041},
          eid = {114501},
        pages = {114501},
          doi = {10.1088/1538-3873/ac9642},
archivePrefix = {arXiv},
       eprint = {2210.02276},
 primaryClass = {astro-ph.IM},
       adsurl = {https://ui.adsabs.harvard.edu/abs/2022PASP..134k4501C}
}

@BOOK{ASTRO2020,
  author    = "{{Astro2020: National Academies of Sciences, Engineering, and Medicine}}",
  title     = "Pathways to Discovery in Astronomy and Astrophysics for the 2020s",
  isbn      = "978-0-309-46734-6",
  doi       = "10.17226/26141",
  url       = "https://nap.nationalacademies.org/catalog/26141/pathways-to-discovery-in-astronomy-and-astrophysics-for-the-2020s",
  year      = 2023,
  publisher = "The National Academies Press",
  address   = "Washington, DC"
}

@INPROCEEDINGS{sho01,
       author = {{Shortridge}, K.},
        title = "{Astronomical Software Strategies}",
    booktitle = {Organizations and Strategies in Astronomy, Volume II},
         year = 2001,
       editor = {{Heck}, Andr{\'e}},
       series = {Astrophysics and Space Science Library},
       volume = {266},
        month = jan,
        pages = {163},
          doi = {10.1007/978-94-010-0666-8_11},
       adsurl = {https://ui.adsabs.harvard.edu/abs/2001ASSL..266..163S}
}


\end{document}